\documentclass[12pt]{article}
\usepackage{epsfig}

\def\beq{\begin{equation}}
\def\eeq#1{\label{#1}\end{equation}}
\def\eeqn{\end{equation}}

\def\beqa{\begin{eqnarray}}
\def\eeqa#1{\label{#1}\end{eqnarray}}
\def\eeqan{\end{eqnarray}}

\let\bar=\overbar

\def\Dslash{\not{\hbox{\kern-4pt $D$}}}
\def\dslash{\not{\hbox{\kern-2pt $\del$}}}

\def\msb{{\bar{\ssstyle M \kern -1pt S}}}

\def\Title#1{\begin{center} {\Large {\bf #1} } \end{center}}

\begin{document}

\Title{Measurements of quantum-correlated $D^{0}\overline{D^{0}}$ decays related to the determination of $\gamma/\phi_3$}
\bigskip\bigskip


\begin{raggedright}  

{\it Jim Libby \index{Libby, J.}\\
Indian Institute Technology Madras \\
Department of Physics\\
Chennai 600036, INDIA}
\bigskip\bigskip
\end{raggedright}

\noindent {\it Proceedings of CKM 2012, the 7th International Workshop on the CKM Unitarity Triangle, University of Cincinnati, USA, 28 September - 2 October 2012} 

\section{Introduction}
One of the main goals of flavour physics is to determine the angle $\gamma/\phi_3$ of the $b-d$ CKM
triangle \cite{CKM}. Aside from being the least well known angle of the unitarity triangle (UT), it can be
determined in tree-level processes that have negligible contributions from beyond the standard model physics,
unlike most other constraints on the UT \cite{FITTERS}. 

Currently all determinations of and constraints on $\gamma/\phi_3$ at tree level come from $B^{-}\to \widetilde{D}^{0}K^{-}$  \cite{GLW}
and related decays involving a $D^{*}$ or $K^{*}$ in the final state.\footnote{Here and throughout
this paper the charge-conjugate state is implied unless otherwise stated.} Here,
$\widetilde{D}^{0}$ is a $D^{0}$ or $\overline{D^0}$ decaying to the same final state. The sensitivity to
$\gamma/\phi_3$ is due to the interference between the decay $B^{-}\to D^{0}K^{-}$ and the colour and
CKM-suppressed decay $B^{-}\to \overline{D^0}K^{-}$. The most precise single measurements of  
$\gamma/\phi_3$ come from decays where $\widetilde{D}^{0}\to K^{0}_{S}h^{+}h^{-}$ \cite{BABAR3,BELLE2}, where $h$ is $\pi$ or $K$. In addition, $CP$ violation has also been observed in the case $\widetilde{D}^{0}\to K^{-}\pi^{+}$ \cite{LHCB_ADS}, which leads to significant constraints on $\gamma/\phi_3$ in combination with other measurements \cite{SHAGGER}. Furthermore, the related final states $K^{-}\pi^{+}\pi^{+}\pi^{-}$, $K^{-}\pi^{+}\pi^{0}$ and $K^{0}_{S}K^{-}\pi^{+}$ \cite{ADS,AS} hold significant sensitivity to $\gamma/\phi_3$. However, all measurements of the aforementioned states depend on parameters related to the decay of the $D^{0}$ meson. Knowledge of the $D$-decay parameters {\it a priori} can greatly improve the determination
$\gamma/\phi_3$. These proceedings summarise the measurements of these $D$-decay parameters made by the CLEO collaboration  \cite{KSHH,LOWREY,K0SKPI} and briefly discuss the potential to improve these measurements using data collected by BESIII.
   
\section{Measurement of the strong-phase parameters of $D^{0}\to K^{0}h^{+}h^{-}$ decays}
The sensitivity to $\gamma/\phi_3$ in $B^{-}\to \widetilde{D}^{0}(K^{0}_{S}h^{+}h^{-})K^{-}$ comes from studying
differences between the $\widetilde{D}^{0}\to K^{0}_{S}h^{+}h^{-}$ Dalitz plot for both $B^{-}$ and $B^{+}$ decays \cite{GIRI,BONDAR1}.
Some measurements of $\gamma/\phi_3$ require a model of the $\widetilde{D}^{0}\to K^{0}_{S}h^{+}h^{-}$ amplitude, which is derived from flavour-tagged samples of $D^{0}\to K^{0}_{S}h^{+}h^{-}$. The assumptions used to
determine the model introduce a systematic uncertainty on $\gamma/\phi_3$ which is estimated to be between
$3^{\circ}$ and $9^{\circ}$ \cite{BABAR3,BELLE2}. This is significantly less than the current statistical
uncertainty but it will be a limiting factor in future measurements \cite{LHCBK0SPIPI,EPLUSEMINUS}. Therefore, it is
desirable to perform the measurement in a model-independent manner. Such a method was proposed in Ref.~
\cite{GIRI} and has been developed significantly by Bondar and Poluektov \cite{BONDAR2}.  The method requires
determining yields in bins of the $\widetilde{D}^{0}\to K^{0}_{S}h^{+}h^{-}$ Dalitz plot for $B^{-}$ and $B^{+}$
decay, which depend on the $B$-decay parameters and two new parameters $c_i$ and $s_i$, which are the
amplitude-weighted averages over the bin of the cosine and sine of the difference in strong-phase difference,
$\Delta\delta_{D}$, between Dalitz-plot points $(m_{-}^{2},m_{+}^{2})$ and $(m_{+}^{2},m_{-}^{2})$. Here $m_{\pm}$
is the invariant-mass of the $K^{0}_{S}h^{\pm}$ pair. It can be shown \cite{BONDAR2,KSHH} that between 80\% to 90\%
of the statistical sensitivity to $\gamma/\phi_3$ of the unbinned method can be obtained by choosing bins 
corresponding to equal intervals of $\Delta\delta_{D}$ according to an amplitude model. Model-independent measurements have now been reported by the Belle \cite{BELLEMODIND} and LHCb \cite{LHCBK0SPIPI} using CLEO-c measurements of the $c_i$ and $s_i$ parameters \cite{KSHH}. 
 
The values of $c_i$ and $s_i$ are measured in quantum-correlated $D^{0}\overline{D^0}$ decays of the
$\psi(3770)$. The $D^{0}\overline{D^0}$ are produced in a $C=-1$ state. Therefore, if one $D$ meson decays to a
$CP$ eigenstate, the other $D$-meson is in the opposite $CP$ eigenstate. The difference between $CP$-even and
$CP$-odd tagged Dalitz plots in each bin is related to the $c_i$ parameters. In addition, the Dalitz plot of 
quantum-correlated events where both $D$-mesons decay to $K^{0}_{S}h^{+}h^{-}$ is sensitive to both $c_i$ and
$s_i$. The strong-phase parameters for the decay $D^{0}\to K^{0}_{L}h^{+}h^{-}$ ($c^{\prime}_{i}$ and
$s^{\prime}_{i}$) are closely related to $c_i$ and $s_i$ such that using decays of the type $K^{0}_{S}h^{+}h^{-}$
$vs.$ $K^{0}_{L}h^{+}h^{-}$ greatly improve the precision on $c_i$ and $s_i$.

The CLEO-c experiment \cite{CLEOC} collected $e^{+}e^{-}\to\psi(3770)\to D\bar{D}$ data corresponding to an
integrated luminosity of 818~pb$^{-1}$. The fact that all particles arise from $D$-meson decay in the final state
leads to both $D$ mesons being reconstructed exclusively with high efficiency and purity. 
For $D^{0}\to K^{0}\pi^{+}\pi^{-}$ ($D^{0}\to K^{0}K^{+}K^{-}$) decay the numbers of CP-tagged and
$K^{0}h^{+}h^{-}$ $vs.$ $K^{0}h^{+}h^{-}$ candidates selected are 1661 and 1674 (219 and 335), respectively. 
A maximum-likelihood fit is performed to the bin yields of the $CP$-tagged and $K^{0}h^{+}h^{-}$ $vs.$
$K^{0}h^{+}h^{-}$ events to extract $c^{(\prime)}_{i}$ and $s^{(\prime)}_{i}$. 
The largest systematic uncertainties arise from the modelling of the background. However,
none of these measurements are systematically limited. 

The current systematic uncertainty on $\gamma/\phi_3$ from the uncertainties on $c_i$ and $s_i$ is $4.3^{\circ}$ \cite{BELLEMODIND}. This is approximately twice that predicted \cite{KSHH}, but the increase is due to the low statistics in some bins with the current $B$ data samples \cite{BELLEMODIND}. Therefore, the uncertainty will reduce once larger data sets are analysed. Further, the uncertainty on $c_i$ and $s_i$ will decrease once measurements are available from BESIII, as discussed in Sec.~\ref{sec:impact}.
 
\section{Measurement of the coherence factor and average strong phase difference}

The rate of decays $B^{-}\to \widetilde{D}^{0}(K^{+}\pi^{-})K^{-}$ is particularly sensitive to $\gamma/\phi_3$
because the two interfering amplitudes are of similar size due to the doubly-Cabibbo suppressed (DCS) $D^{0}$ decay
coming from the favoured $B^{-}$ amplitude \cite{ADS}. The rate depends not only on $\gamma/\phi_3$ but on the
strong-phase difference between the Cabibbo-favoured and DCS $\widetilde{D}^{0}\to K^{+}\pi^{-}$ decays. The
measurement of this parameter by the CLEO collaboration is described elsewhere \cite{TQCAII}.

Via the same mechanism there is potential sensitivity to $\gamma/\phi_3$ from $B^{-}\to \widetilde{D}^{0}K^{-}$,
where  $\widetilde{D}^{0} \to K^{+}\pi^{-}\pi^{0}$ or $\widetilde{D}^{0} \to K^{+}\pi^{-}\pi^{-}\pi^{+}$ \cite{AS}.
These modes have significantly larger branching fractions than $\widetilde{D}^{0}\to K^{+}\pi^{-}$ \cite{PDG}. However, the dynamics are more complicated because there is variation of the strong-phase difference over the
multi-body phase-space. This leads to the introduction of a new parameter referred to as the coherence factor $R_F$
($F=K\pi\pi^{0}$ or $K3\pi$), which multiplies the interference term sensitive to $\gamma/\phi_3$. The value of
$R_F$ can vary between zero and one. If there is only a single intermediate resonance or a few non-interfering
resonances the coherence factor will be close to one and the decay will behave just like $\widetilde{D}^{0}\to
K^{+}\pi^{-}$. If there are many overlapping intermediate resonances the coherence factor will tend toward zero,
limiting the sensitivity to $\gamma/\phi_3$; the sensitivity can be recovered by binning the phase space appropriately and computing the coherence factor and average strong phase within the bins. However, even if there is limited sensitivity to $\gamma/\phi_3$ when
$R\sim 0$ there is enhanced sensitivity to the magnitude of the amplitude ratio between the $B^{-}\to D^{0}K^{-}$
and $B^{-}\to \overline{D^0}K^{-}$ decays; improved knowledge of this parameter will then lead to better overall
sensitivity to $\gamma/\phi_3$ in a combined fit to all $B^{-}\to \widetilde{D}^{0}K^{-}$ decays \cite{SHAGGER}.

\begin{figure}[htb]
\begin{center}
\begin{tabular}{cc}
\epsfig{file=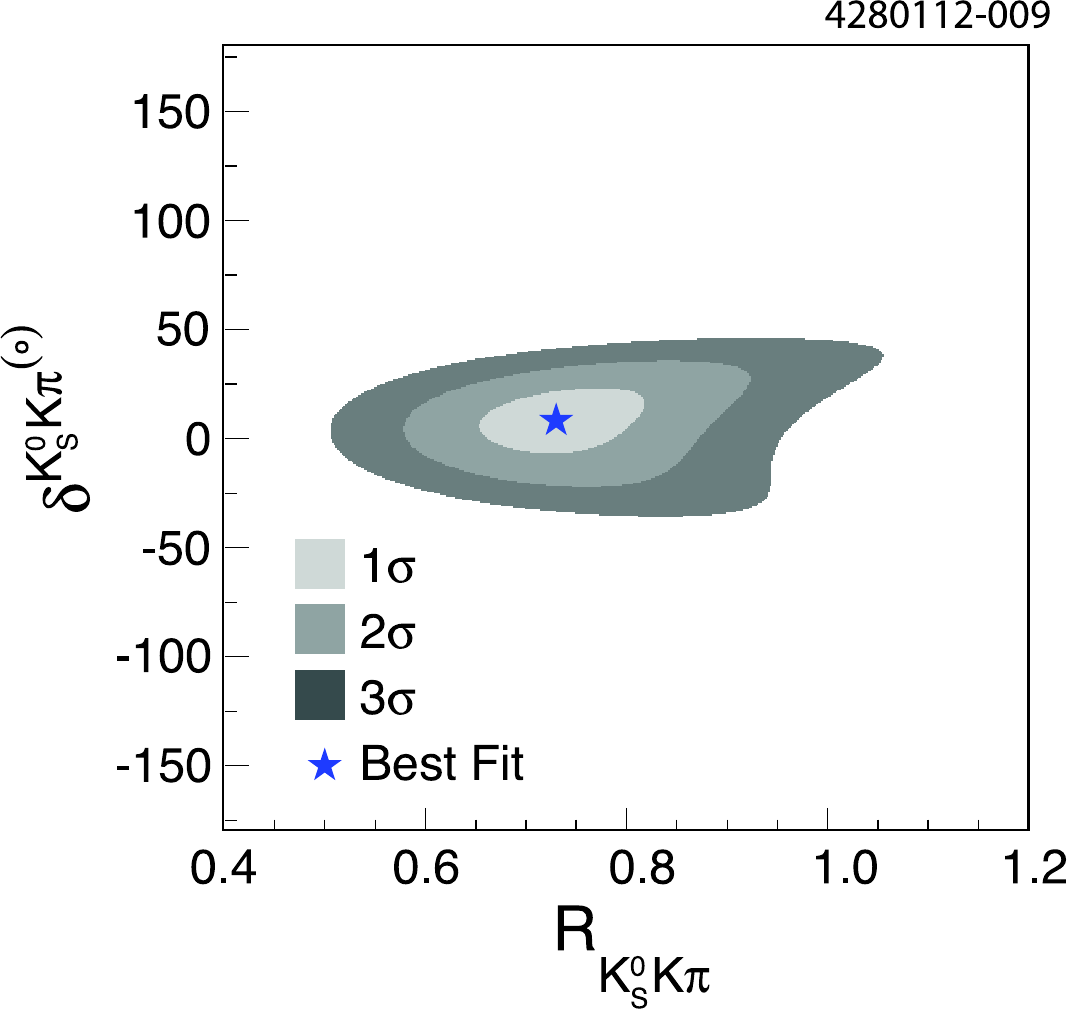,width=0.4\columnwidth} & 
\epsfig{file=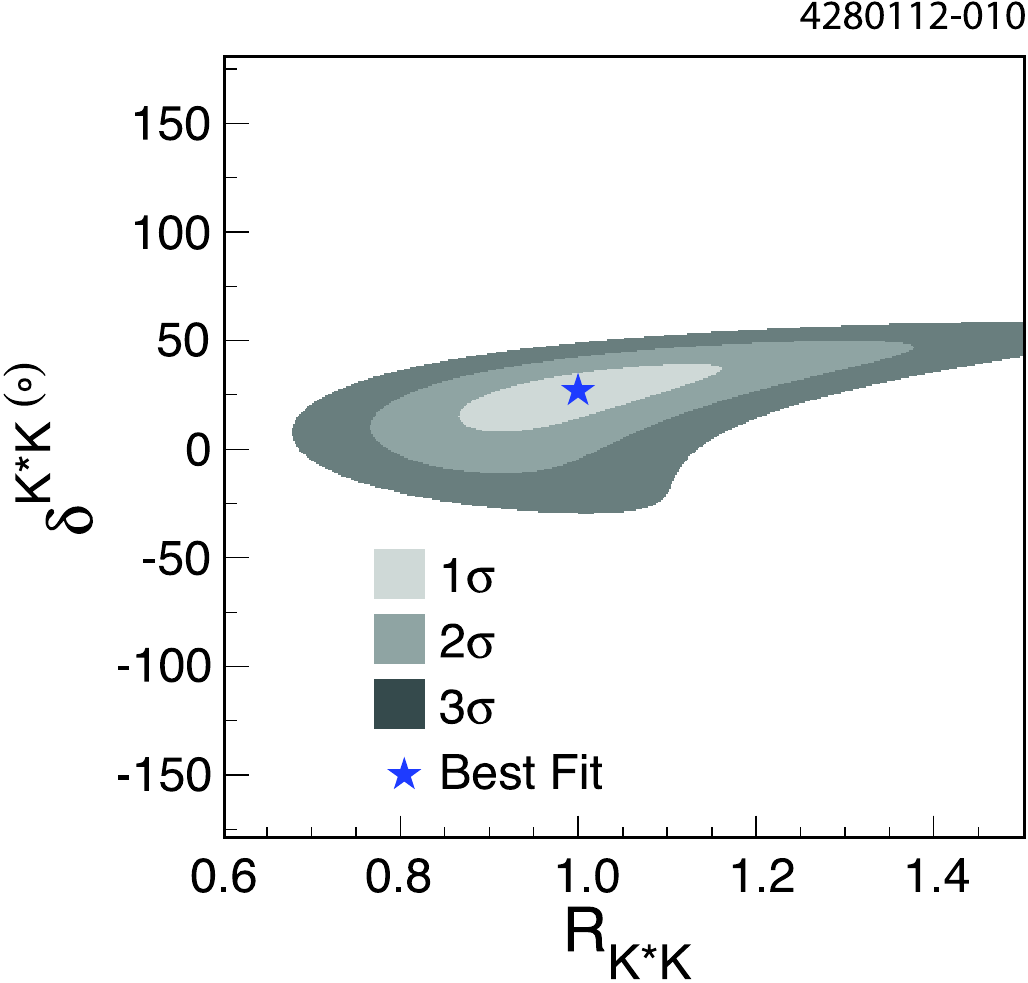,width=0.4\columnwidth} 
\end{tabular}
\caption{The $1\sigma$, $2\sigma$, and $3\sigma$ allowed regions of (left) ($R_{K^{0}_{S}K\pi}$,$\delta_{D}^{K^{0}_{S}K\pi}$) and 
(right) ($R_{K^{*}K}$,$\delta_{D}^{K^{*}K}$)
parameter space.}\label{fig:coherence}
\end{center}
\end{figure}

The values of $R_F$ and the average-strong phase difference $\delta_{D}^{F}$ have been measured by
CLEO-c \cite{LOWREY}. Sensitivity comes from the quantum-correlated $D^{0}\overline{D^0}$ events with $F$ tagged by
either $CP$ eigenstates or $K^{-}\pi^{+}$, $K^{-}\pi^{+}\pi^{0}$, and $K^{-}\pi^{+}\pi^{+}\pi^{-}$, where the tag
kaon charge is the same as the signal. A $\chi^{2}$ fit to the yields gives: $R_{K\pi\pi^{0}}=0.84\pm 0.07$,
$\delta_{D}^{K\pi\pi^{0}}=(227^{+14}_{-17})^{\circ}$, $R_{K3\pi}=0.33^{+0.26}_{-0.23}$, and
$\delta_{D}^{K3\pi}=(114^{+26}_{-23})^{\circ}$. 

Recently measurements of the coherence factor have been made in analogous fashion in the decay $D^{0}\to K^{0}_{S}K^{-}\pi^{+}$ \cite{K0SKPI}. The branching fraction for this decay is smaller than the other modes discussed but the interference is enhanced as the amplitude of the suppressed decay is around 60\% that of the favoured mode. The coherence factor has been measured over the whole $K^{0}_{S}K^{-}\pi^{+}$ phase space and in the region around the $K^{*+}\to K^{0}_{S}\pi^{+}$ resonance.  The sensitivity has been improved by including events tagged by $K^{0}_{S(L)}\pi^{+}\pi^{-}$ and including the measured values of $c_i^{(\prime)}$ and $s_i^{(\prime)}$ \cite{KSHH}. Figure~\ref{fig:coherence} shows the $1\sigma$, $2\sigma$, and $3\sigma$ regions of $(R,\delta_{D})$ parameter space for both kinematic regions; $D^{0}\to
K^{*+}K^{-}$  is observed to be fully coherent. It has been estimated that using 9000 reconstructed $B^{-}\to \widetilde{D}^{0}(K^{*\pm}K^{\mp})K^{-}$ events the uncertainty on $\gamma/\phi_3$ will be $7^{\circ}$ \cite{K0SKPI}. 

\section{Conclusion}
\label{sec:impact}

The measurements of quantum-correlated $D$ decay parameters related to $\gamma/\phi_3$ are being used by {\it BABAR} \cite{BABARKPIPI0}, Belle \cite{BELLEMODIND} and LHCb \cite{LHCBK0SPIPI,SHAGGER}. CLEO-c will update the $K^{-}\pi^{+}\pi^{0}$ and $K^{-}\pi^{+}\pi^{+}\pi^{-}$ coherence factor using events tagged by $K^{0}_{S,L}\pi^{+}\pi^{-}$ to improve the precision. In addition,
CLEO-c has reported an amplitude model for $D^{0}\to K^{+}K^{-}\pi^{+}\pi^{-}$ decays \cite{KKPIPI}, which exploits quantum-correlated, as well as flavour-tagged, samples to better determine the phases. This $D$ final state can also be used to determine $\gamma/\phi_3$ in $B^{-}\to\widetilde{D^{0}}K^{-}$ decays; it has been estimated that 2000 $B$ decays would give a precision of $11^{\circ}$ on $\gamma/\phi_3$ using a model-dependent analysis \cite{KKPIPI}.

The future of quantum-correlated measurements lies with BESIII \cite{BESIII} which has collected a $2.9~\mathrm{fb}^{-1}$ data  set at the $\psi(3770)$. It is foreseen that the size of this data set will be increased to $10~\mathrm{fb}^{-1}$ by the end of 2015 \cite{ROY}.
Measurements of $D$-decay parameters with such a data set should lead to uncertainties on $D$ parameters that will not limit the precision on $\gamma/\phi_3$ even at future $e^{+}e^{-}$ facilities and an upgraded LHCb. First measurements of quantum-correlated parameters at BESIII are underway \cite{ROY}. It should also be noted that the measurements of $c_i$ and $s_i$ also allow a model-independent determination of charm-mixing and $CP$-violation parameters \cite{BPV}. It has been shown that BESIII measurements will again be sufficient to prevent the mixing-parameter precision being systematically limited at future facilities \cite{TW}. Finally there are as yet unexplored modes, $\pi^{+}\pi^{-}\pi^{0}$ and $K^{0}_{S}\pi^{+}\pi^{-}\pi^{0}$, which may yield significant sensitivity to $\gamma/\phi_3$; quantum-correlated measurements of $D$ parameters will be an important component of exploiting their potential.

\section*{Acknowledgments}

I would like to thank Roy Briere, Hai-bo Li and Hajime Muramatsu for providing information about the quantum-correlated programme at BESIII.


\end{document}